\documentclass[useAMS,usenatbib,preprint,authoryear,12pt]{elsarticle}

 \usepackage{graphics}
 \usepackage{graphicx}
\usepackage{epsfig}

\usepackage{amssymb}

\journal{New Astronomy}

\begin{document}

\begin{frontmatter}



\title{Do Damped and Sub-damped Lyman-alpha Absorbers Arise in Galaxies of Different Masses?}


\author[1]{Varsha P. Kulkarni}
\address[1]{Department of Physics and Astronomy, University of South Carolina,
    Columbia, SC 29208, USA}
    
\author[2]{Pushpa Khare}
\address[2]{Department of Physics, Utkal University, Bhubaneswar, 751004, India}

 \author[1]{Debopam Som}

 \author[3]{Joseph Meiring}
\address[3]{Department of Physics \& Astronomy, University of Louisville, Louisville, KY 40292, USA}

    \author[4]{Donald G. York}
\address[4]{Department of Astronomy \& Astrophysics, University of Chicago, Chicago, IL 60637, USA}

    \author[5]{Celine P\'eroux}
\address[5]{Laboratoire dÕAstrophysique de Marseille, OAMP, UniversiteAix-Marseill\'e \& CNRS, 13388 Marseille cedex 13, France }

    \author[3]{James T. Lauroesch}

\begin{abstract}
We consider the questions of whether the damped Lyman-alpha (DLA) and sub-DLA absorbers in quasar spectra differ intrinsically in metallicity,  and whether they could arise in galaxies of different masses.  Using the recent measurements of the robust 
metallicity indicators Zn and S in DLAs and sub-DLAs, we confirm that sub-DLAs 
have higher mean metallicities than DLAs, especially at $z \lesssim 2$. We find that the intercept of the metallicity-redshift relation derived from Zn and S is higher than that derived from Fe by 0.5-0.6 dex. We also show that, while there is a correlation between the metallicity and the 
rest equivalent width of Mg II $\lambda 2796$ or Fe II $\lambda 2599$ for DLAs, no  
correlation is seen for sub-DLAs. Given this, and the similar Mg~II or Fe~II  selection criteria employed in the discovery of both types of systems at lower redshifts, the difference between
metallicities of DLAs and sub-DLAs appears to be real and not an artefact of selection.  This conclusion is supported by 
our simulations of Mg II $\lambda 2796$ and Fe II $\lambda 2599$ lines for a wide range of physical conditions. On examining the velocity spreads of 
the absorbers, we find that sub-DLAs show somewhat higher mean and median velocity spreads ($\Delta v$), and an excess of systems with  $\Delta v > 150$ km s$^{-1}$, than DLAs. Compared to DLAs, the [Mn/Fe] vs. [Zn/H] trend for sub-DLAs appears to be steeper and closer to the trend for Galactic bulge and thick disk stars, possibly suggesting different stellar populations. The absorber data appear to be consistent with galaxy down-sizing. The data are also consistent with the relative number densities of low-mass and high-mass galaxies. It is thus plausible that sub-DLAs arise in more massive galaxies on average than DLAs.

\end{abstract}


\end{frontmatter}


\section{Introduction}
\label{1}
The evolutionary history of galaxies is written in part in the interstellar abundances of the chemical 
elements. Deciphering this history from observations of distant galaxies is an important 
aspect of understanding the various processes responsible for evolution of galaxies. Besides the 
stellar mass function, yields, supernova rates etc., the metallicity of a galaxy is also 
governed by the depth of its gravitational potential well, which determines the ability to retain the metals produced 
from star formation. Indeed, the interstellar metallicity (as determined from nebular emission lines) is found to be correlated tightly with the stellar mass of the galaxy  at $z\sim 0$ and also at $0.4 < z < 3$ (e.g., Tremonti et al. 2004; Savaglio et al.  2005; Erb et al.  2006). 

Quasar absorption lines provide a powerful complementary tool for studying distant galaxies. 
The damped Lyman-alpha (DLA; log $N_{\rm H I} \ge 20.3$) and sub-damped Lyman-alpha 
(sub-DLA; $19.0 \le$ log $N_{\rm H I} < 20.3$) absorbers are the primary neutral gas reservoir 
at $0 < z < 5$  (e.g.,  Storrie-Lombardi \& Wolfe 2000; P\'eroux et al. 2005; Prochaska et al. 2005), and offer the most precise element abundance measurements in distant galaxies. 

As per most chemical evolution models, the mean 
interstellar metallicity of galaxies should reach a near-solar value at low redshift, as a result of 
progressive generations of star formation. 
Surprisingly, most DLAs observed at $0.1 < z < 3.9$ are metal-poor (e.g., Kulkarni et al 2005, 2007, and references therein).  By contrast, a 
substantial fraction of the sub-DLAs observed at $0.6 < z < 1.5$ are metal-rich--some even super-solar 
 (e.g., Kulkarni et al. 2007; Meiring et al. 2007, 2008, 2009a, 2009b; P\'eroux et al. 2006, 2008; and references therein).  
Based on the observed mass-metallicity relation for galaxies,  Khare et al. (2007) suggested that the sub-DLAs may be more massive compared to DLAs. 
Here we examine various selection effects that could potentially be responsible 
for this conclusion with an updated sample of DLA/sub-DLA metallicities, and compare other properties of DLAs and sub-DLAs.  We use 
the nearly undepleted elements Zn and S, which makes our analysis ``cleaner'' and more conservative than studies based on depleted elements such as Fe.  This also allows us to make meaningful comparisons of the observed DLA/sub-DLA metallicity evolution with known trends for galaxies.

\section{Sample Description, Observed Trends, and Metallicity Indicator}
\label{2}
\subsection {Sample Description} 
Comparison of the metallicity evolution of DLA and sub-DLA galaxies was made in Kulkarni et al. (2007) based on Zn data for 119 DLAs and 30 sub-DLAs. In this paper, we examine this issue further using Zn and S for 154 DLAs and 58 sub-DLAs. 
The two-fold increase in the number of sub-DLAs since 2007 has 
come primarily from our recent measurements (20 systems from Meiring et al.  2008; P\'eroux et al. 2008; Meiring et al. 2009a), complemented by a few measurements from the literature 
(Nestor et al. 2008, Quast et al. 2008, Noterdaeme et al. 2008, Dessauges-Zavadsky et al. 2009).  
To avoid any prior bias toward high or low metallicity, we exclude systems which were specifically observed because of hints that they are rich in metals or molecules. (Thus, we exclude two metal-rich sub-DLAs from Prochaska et al. 2006, one metal-rich sub-DLA from Srianand et al. 2008, and the metal-strong DLAs from Herbert-Fort et al. 2006. In any case, we note Kulkarni et al. 2007 showed that adding the metal-strong systems did not change the conclusions about relative evolutions of DLAs and sub-DLAs.)

\subsection{Observed Trends}

Fig. 1 shows the $N_{\rm H I}$-weighted mean metallicity vs. look-back time relation for the 154 DLAs and 58 sub-DLAs in our sample, calculated 
using the procedures outlined in Kulkarni \& Fall (2002). Upper limits on Zn have been treated with survival analysis. 
Also shown for reference are the predictions for metallicity evolution in two theoretical models. The light dot-dashed curve shows the mean interstellar metallicity  in the chemical evolution model 
of Pei et al. (1999) with the optimum fit for the cosmic infrared background intensity. This model also 
uses observational constraints from optical galaxy surveys and the comoving H I density of DLAs  to calculate the coupled global evolution of stellar, gaseous, and metal content of galaxies. 
The light dot-double-dashed curve shows the mean metallicity of cold interstellar gas in a semianalytic model of galaxy formation in the cold dark matter merging hierarchy by Somerville et al. (2001). This model, referred to as the ``collisional starburst model'' by Somerville et al., assumes star formation in bursts triggered by galaxy mergers in addition to a quiescent star formation at constant efficiency.  

The sub-DLA global mean metallicity appears to be higher than that of DLAs, reaching a near-solar value at
low $z$, consistent with the models (see also York et al. 2006; Prochaska et al. 2006).  
The linear regression slope for the sub-DLA $N_{\rm H I}$-weighted mean metallicity vs. redshift 
data ($-0.46 \pm 0.18$) appears to be steeper than that for DLAs ($-0.22 \pm 0.07$). The 
slope for sub-DLAs has larger uncertainties due to the smaller sample size. The corresponding 
linear regression estimates of the intercepts, i.e. expected metallicities at $z=0$, are $0.18 \pm 0.29$ for sub-DLAs and $-0.63 \pm 0.16$ for DLAs, which differ at 2.4 $\sigma$ level.  The bold curves in Fig. 1 show the best fits to the DLA and sub-DLA data (plotted in terms of look-back time). The results do not change substantially if the metallicities are unweighted (a slope of $-0.47 \pm 0.16$ and an intercept of $0.22 \pm 0.24$ for sub-DLAs, and a slope of $-0.25 \pm 0.05$, intercept of $-0.47 \pm 0.11$ for DLAs.) 
Below some redshift around 2, the mean metallicity of sub-DLAs exceeds that of DLAs. Determining exactly where that happens (e.g. at $z< 1.7$ or $z<2.2$) will require a larger database. 

\subsection{The Choice of the Metallicity Indicator}

We have used measurements of Zn or S, since these nearly undepleted elements offer the most direct ``dust-free'' metallicity indicators. Furthermore  S and Zn are seen to track each other well (within $\lesssim  0.2$ dex) in  
most absorbers, as seen in Fig. 2, which shows [S/Zn] vs. [Zn/H] for absorbers with detections of both S and Zn (see also Nissen et al. 2004, 2007).
Fe, used in some previous works, is not a reliable metallicity indicator since it is severely depleted on interstellar dust, so that its gas-phase abundance does not give its total abundance  
(see, e.g., Miller et al. 2007; Jenkins 2009). Moreover, depletions 
correlate with metallicity: [Zn/Fe] $\sim 0.2$ for [Zn/H] $\sim -1.5$ and  [Zn/Fe] $\sim 1$ for [Zn/H] $\sim 0$ and can show scatter (e.g., Ledoux et al. 2003; Meiring et al. 2006). Thus, [Fe/H] can seriously underestimate the true metallicity, especially at the high-metallicity end, and may suggest a less steep evolution than the true trend. It is interesting that the slope for sub-DLAs from unweighted Fe/H ($-0.41 \pm 0.07$; Dessagues-Zavadsky et al. 2009) is close to our value $-0.47 \pm 0.16$ based on unweighted X/H (where X = Zn or S), but the 
intercepts based on unweighted Fe/H ($-0.26 \pm 0.13$ and $-1.03 \pm 0.09$ for sub-DLAs and DLAs, respectively) are substantially lower than those based on unweighted X/H  ($0.22 \pm 0.24$ and $-0.47 \pm 0.11$ for sub-DLAs and DLAs, respectively).  
Thus, using Fe systematically underestimates the intercept of the metallicity-redshift trend by 0.5-0.6 dex for both DLAs and sub-DLAs. 

\section{Is the Metallicity Difference between DLAs and Sub-DLAs a Selection Effect?}
\label{3}

We now consider whether the higher metallicity of sub-DLAs 
could be an artefact. Ionisation effects 
and dust selection have been ruled out before as the cause of the 
difference between DLA and sub-DLA metallicities (e.g., Khare et al. 2007).  We mainly consider here the possiblity of a selection effect tied to the presence of strong Mg II $\lambda$ 2796 and Fe II $\lambda 2599$ lines.

\subsection{Is there an Mg II Selection Bias?}
The sample of sub-DLAs studied by our team over the past few years (Meiring et al. 2009b and references therein) was chosen from the low-$z$ sub-DLAs with measured $N_{\rm H I}$ 
values. As emphasized in our previous papers, the availability of $N_{\rm H I}$ was the primary selection criterion, not the strength of any metal lines. However, the HST sample of DLAs and sub-DLAs that S. Rao and collaborators chose to measure $N_{\rm H I}$ are systems with  rest-frame equivalent widths for the Mg II $\lambda 2796$ line exceeding  0.5 {\AA} ($W_{2796} > 0.5$ \AA).  Below we discuss whether this introduces a bias toward more metal-rich systems in the sub-DLAs.  
Simply put, the argument is that a system with large $W_{2796}$ would have 
large $N_{\rm Mg II}$, hence  a large $N_{\rm Zn II}$, and hence a higher metallicity for 
low $N_{\rm H I}$. There are several problems with this argument:

{\bf Mg II Line Saturation:} 
The Mg II $\lambda 2796$ line is almost always highly saturated in DLAs and 
sub-DLAs. The systems with larger $W_{2796}$ have 
larger velocity spreads, not larger Mg II column densities. 
Thus, selecting systems with larger $W_{2796}$ does not necessarily translate into 
selecting systems with larger $N_{\rm Mg II}$ and hence larger $N_{\rm Zn II}$. 

{\bf Lack of Correlation between Mg II Strength and Metallicity:} If there were an Mg II selection bias, one would expect to see a  positive correlation between 
metallicity and  $W_{2796}$ for sub-DLAs. However, no correlation is seen between 
[X/H] (where X = Zn or S) and $W_{2796}$ for sub-DLAs (see Fig. 3a). Using the generalized Kendall's $\tau$-correlation coefficient (including survival analysis to treat upper limits), we get $\tau = -0.070$, and a 2-sided probability of 0.670 for no correlation (see Table 1). Even [Fe/H] and $W_{2796}$ do not correlate for sub-DLAs, with generalized Kendall's $\tau = 0.128$ and a 
2-sided probability of 0.541 for no correlation for points from Fig. 3 of Dessagues-Zavadsky et al. (2009).   
By contrast, a significant correlation between Mg II strength and [X/H] is seen 
for DLAs (Fig. 3b), with a generalized Kendall's $\tau$ = 0.819, and a probability of $< 0.0001$ for no correlation (Table 1).  Even [Fe/H] shows a significant correlation with $W_{2796}$ for DLAs (see also Fig. 1 of Murphy et al. 2007).  Thus, the sub-DLAs appear to be far less likely than the DLAs to show an Mg II-selection bias toward higher metallicity. The smaller correlation in the sub-DLA sample is apparent even 
looking at just the detections plotted in Figs. 3(a) and 3(b). 
It is possible that this difference between DLAs and sub-DLAs is caused partly by the smaller number of sub-DLAs plotted in Fig. 3(a), but the number of DLAs  in Fig. 3(b) is not much larger, and the DLAs appear to show a correlation across most of the range of $W_{2796}$. 

{\bf Lack of Sensitivity Problem:} Kulkarni et al. (2007) used simulations of 
the Mg II $\lambda 2796$ line with a variety of velocity component structures to estimate the 
minimum measurable [Mg/H] in the observed sub-DLA  sample. 
These simulations showed that, for the choice of $W_{2796} \ge 0.5$ \AA \, 
one can detect systems with [Mg/H] well below $-1.5$ for log $N_{\rm H I} = 19.7$. 
Thus, choosing systems with $W_{2796} \ge 0.5$ \AA \, cannot explain the fact that 
$\sim 50 \%$ of the low-$z$ sub-DLAs with Zn or S detections have [X/H] $> -0.5$, and only $\sim 10 \%$ systems with Zn detections have [X/H] $< -1.0$. 

{\bf Representativeness of Sub-DLA Sample:} One may wonder whether the sub-DLAs with $W_{2796} > 0.5$ {\AA} are unrepresentative of the sub-DLA population in terms of metallicity. However, for this to produce a bias in the current measurements, the number distribution of sub-DLAs would have to be skewed to have many more systems with $W_{2796} < 0.5$ {\AA} than with $W_{2796} > 0.5$ {\AA}, and those objects would have to be more metal-poor than what is found now for sub-DLAs. This does not seem probable. More important, such a scenario would imply a large missing fraction of sub-DLAs that would be difficult to reconcile with current measurements of the number density $dn/dz$ and the column density distribution $f(N_{\rm H I})$ which are consistent with both DLAs on the one hand and Lyman limit systems on the other hand. 

{\bf Metallicity correlation with $E(B-V)$:} 
York et al. (2006, hereafter Y06), using $> 800$ SDSS Mg II systems and assuming a constant dust-to-gas ratio, 
found an anti-correlation between quasar reddening $E(B-V)$ and metallicity.  This trend becomes more pronounced for a metallicity-dependent dust-to-gas ratio (Khare et al. 2007).  Y06 also showed that strong  $W_{2796}$ can occur for all values of $E(B-V)$, which implies that systems with strong Mg~II 
lines could correspond to high or low metallicity, i.e. 
using systems with large $W_{2796}$ does not bias one toward selecting only metal-rich 
systems. 

We thus conclude that the higher sub-DLA metallicities are not caused by an Mg II selection bias. 

 \subsection{Is there an Fe II Selection Bias?} 

Another potential bias could come from the fact that the low-$z$ DLA and sub-DLA absorbers chosen for the HST UV spectroscopy by Rao et al. also have relatively strong Fe II $\lambda 2599$ lines 
(rest frame equivalent widths $W_{2599} > 0.5$ {\AA}).  The squares and circles in Figs. 4(a) and 4(b)  show the observed metallicity vs. $W_{2599}$ for the sub-DLAs and DLAs, respectively,
in our sample for which $W_{2599}$ measurements are available. 
A correlation between the observed metallicity and the strength of Fe II $\lambda 2599$  is seen for DLAs, but not for sub-DLAs (Table 2). This shows that the sub-DLA data have no systematic bias toward high metallicity systems. 
 
To examine this issue further, we explored the variation of $W_{2599}$ with 
metallicity by simulating Fe II $\lambda 2599$ lines 
in DLAs and sub-DLAs considering a range of physical conditions of the absorbing gas. The H I column densities were chosen to resemble the ranges observed in our sample: for DLAs, randomly in the range 20.3-22.0, and for sub-DLAs, such that 3/4 of all the randomly selected values lie in the range 19.7-20.3, and 1/4 in the range 19.0-19.7. For each selected value of 
N$_{\rm H I}$, the metallicity [X/H] was selected randomly between -2.5 and +1.0, and 
the depletion factor for Fe was selected randomly within a range of -0.6 to -1.2 dex. [This is a fairly conservative Fe depletion range in warm Galactic ISM.  The low-$z$ sub-DLAs with Zn detections show [Fe/Zn] in the range of -0.1 to -1.6 dex, most between -0.4 and -0.9 dex (e.g., Meiring et al. 2009b).]  The number of velocity components in each artificial profile was 
chosen randomly between 3 and 15 having random velocity values within a range constrained by the selected number of components (the maximum range being -375 to 375 km s$^{-1}$ for 15 components). The components were assigned random Doppler $b$ parameters 
ranging between 3 and 20 km s$^{-1}$. The 
equivalent width of each resulting Fe II $\lambda 2599$ line was measured. Finally, the results for the 30,000 artificial lines generated in this way were binned over metallicity into 0.25 dex bins. The diamonds in Figs. 4(a) and 4(b) show the median equivalent width in each bin plotted vs. the central metallicity of the bin. The error bars on the median equivalent widths in each metallicity bin show the $\pm 1 \, \sigma$ variations among the equivalent widths generated from the 
simulations with metallicities lying in that bin. Over $\sim 2/3$ of 
the observed detections lie within the $\pm 1 \sigma$ error bars, as may be expected. 
 
Figs.  4(a) and 4(b) suggest that one may expect a gentle trend of gradually increasing $W_{2599}$ with increasing metallicity. Indeed, the observed DLAs do show a correlation roughly resembling the median simulated trend, except that no supersolar DLAs are seen. By contrast, although most sub-DLA detections lie within the $\pm 1 \sigma$ range 
of the simulated trend, they show no resemblance to the median simulated trend. The lack of correlation among the sub-DLA points may be a result of the small sub-DLA sample. The cluster of points lying to the right of the simulated trend in Fig. 4(a) suggest that the depletions in some of the  observed sub-DLAs may be less severe than assumed above, or the velocity components may be more widely spread. 
In any case, it is clear from Fig. 4(a) that a sample with $W_{2599} > 0.5$ {\AA} can have 
sub-DLAs with [X/H] as low as -1.1 even for the relatively large 
depletion factors assumed above (as low as -1.3 within the 2 $\sigma$ range). All of the observed metallicities for sub-DLAs with detections are considerably higher than that.  Indeed, Fe II selection does not  explain why several sub-DLAs have super-solar metallicity, while DLAs do not. 

We conclude that the higher mean sub-DLA metallicities compared to DLAs do not appear to arise from an Fe II selection bias. This conclusion is consistent with the lack of Mg II selection bias (section 3.1), since Fe II $\lambda$ 2599 is fairly close in strength to Mg II $\lambda$ 2796 (log [$\lambda \times f] = 2.793$ for Fe II $\lambda$ 2599 and 3.236 for Mg II $\lambda$ 2796).

\subsection{\bf{Do abundance gradients play a role?} }
It has been suggested that DLAs could be metal-poor because gas cross-sections are larger in the less developed outskirts of galaxies (e.g., Chen et al. 2005; Ellison et al. 2005; Zwaan et al. 2005). 
However, abundance gradients are relatively weak in the Galactic disk, 
e.g., an [Fe/H] gradient of -0.052 dex kpc$^{-1}$ for 5-17 kpc and -0.012 dex kpc$^ 
{-1}$ for 10-15 kpc (Lemasle et al. 2008). 
The SDSS has shown that the stellar metalicity gradient at the solar circle is 
low, $\lesssim 0.01$ dex kpc$^{-1}$ (Ivezi\'c  et al. 2008). 
Metallicity gradients are also weak in nearby spirals, 
$\sim$ -0.09 dex kpc$^{-1}$ (Ferguson et al. 1998), and it is unclear whether the gradients exist in distant galaxies.  In fact, given that H I column densities are lower in sight-lines through outer parts of local galaxies, one may expect 
sub-DLAs to have lower abundances than DLAs.  Thus abundance gradients cannot explain the large difference between the metallicities of DLAs and sub-DLAs.  

\smallskip
\noindent { \bf{How well-determined are the H I column densities?} }
As discussed by Meiring et al. (2007, 2008, 2009a--see appendices) and by Ellison, Murphy, \& Dessauges-Zavadsky et al. (2009a),  the $N_{\rm H I}$ values of sub-DLAs inferred from single-component fits to low-resolution spectra can be overestimates by 0.1-0.3 dex. 
(If there is a wide spread of components and $N_{\rm H I}$ is low, a single-component H I  fit forces the damping profile to contain more H I that is actually there. 
The effect is not important for DLAs.) Thus, the true metallicities of sub-DLAs can be 
higher by 0.1-0.3 dex, making the difference between DLAs and sub-DLAs even larger.  

\smallskip
We conclude that sub-DLAs are, on average, intrinsically more metal-rich than DLAs. 
Near-solar or super-solar abundances have also been estimated in some Lyman-limit systems and 
even some systems with log $N_{\rm H I}  < 16$ (e.g., Ding et al. 2003; Maseiro et al. 
2005; Jenkins et al. 2005; Misawa et al. 2008). These studies suggest (despite the dependence on details of photoionisation calculations) the intriguing possibility of the $N_{\rm H I}$-metallicity anti-correlation 
extending over $\sim 6$ orders of magnitude in $N_{\rm H I}$.

\section{The Velocity Spreads of DLAs and Sub-DLAs}
\label{4}

One may wonder whether the metallicity difference between DLAs and sub-DLAs is related to   
kinematic differences. We therefore compare the velocity spreads of the absorbers.  

\subsection{Velocity Spread Distributions}
The velocity spread $\Delta v$ for DLAs and sub-DLAs is usually derived by integrating 
the apparent optical depth profile over the extent of a metal line, and calculating the velocity width 
enclosing the inner $90 \%$ of the absorption.  While the overall range of $\Delta v$ 
is similar for our DLA and sub-DLA samples, the mean velocity spread is 
125.5 km s$^{-1}$ for sub-DLAs and 103.0 km s$^{-1}$ for DLAs. The median values 
are 95.3 km s$^{-1}$ for sub-DLAs and  80.0 km s$^{-1}$ for DLAs. Thus sub-DLAs have somewhat 
larger velocity spreads  on average than DLAs. 
Furthermore, the fraction of systems with $\Delta v > 150$ km s$^{-1}$ relative to 
those with $\Delta v < 50$ km s$^{-1}$ is greater among sub-DLAs than DLAs 
(see Table 3). Assuming Poissonian error bars on the number of systems detected in either  velocity range, sub-DLAs show a 2.0 $\sigma$ excess of systems with $\Delta v > 150$ km s$^{-1}$ as compared to DLAs. Of course, larger samples of both DLAs and sub-DLAs are needed to 
verify the difference. 

\subsection{Velocity-Metallicity Correlations}
Correlations between velocity spreads and metallicities might exist for physical reasons. For instance, lower mass galaxies might develop higher wind velocities in outflows. Ledoux et al. (2006) reported a velocity-metallicity correlation in a sample of 62 DLAs combined with 8 high-$N_{\rm H I}$ sub-DLAs, and suggested that it reflects a mass-metallicity relation. 
Figs. 5(a) and 5(b) show velocity spread vs. metallicity for sub-DLAs and DLAs in our sample that have 
velocity measurements available.  A generalized Kendall's $\tau$ correlation test reveals that the correlation of [X/H] with $\Delta v$ is significant for DLAs, but not for sub-DLAs  (Table
4). We find that the correlation between [Fe/H] and $\Delta v$ from Fig. 2 of Dessagues-Zavadsky et al. (2009) is a little stronger than the trend between [X/H] and $\Delta v$, but still marginal for sub-DLAs. 
A weaker correlation with $\Delta v$ for [X/H] than for [Fe/H] could be 
partly due to the lack of a significant number of systems with low [Zn/H], or it could be caused by 
the Routly-Spitzer effect (Routly \& Spitzer 1952): systems with larger $\Delta v$ may have more supernova-driven shocks capable of destroying the interstellar dust grains, resulting in lower depletion 
of iron, and hence higher [Fe/H]. 

We note that there is no unique or perfect way to measure the velocity width, which is very sensitive to the S/N and the spectral resolution. Furthermore,  
for both DLAs and sub-DLAs, the mean or median $\Delta v$ derived from Mg~II $\lambda 2796$ are about a factor of 2 larger than those derived from the weaker lines (Table 3). As an extreme example, we note that the sub-DLA toward Q0515-4414 has 
$\Delta v$ = 573 km s$^{-1}$ from the widely spread Mg II $\lambda$ 2796 line, but only 
94 km s$^{-1}$ from the weaker lines! 
Indeed, if $\Delta v$ from Mg II $\lambda 2796$ is used, the 
correlation with metallicity [X/H] becomes less robust for DLAs as well as sub-DLAs (see Table 4), although this could 
be partly due to the much smaller number of systems with [X/H] that also have 
$\Delta v$ measured from Mg II lines. Thus we caution against strong claims and interpretations 
related to the velocity spreads of DLAs and sub-DLAs. But we note that the larger velocity 
spread in Mg II  $\lambda 2796$ indicates the presence of high-velocity components that are hard to see in the weaker lines, possibly suggesting a substantial amount of halo gas or high velocity clouds. (Mg II in the high velocity clouds in the Milky Way can be inferred from other low ions such as Ca II and Si II [e.g., Wakker et al. 2007; Shull et al. 2009].)

\section{\bf Can Metal-rich Sub-DLAs be more Massive than Metal-poor DLAs?: Comparison with Galaxy Data}
\label{5}

Based on the higher mean metallicity of sub-DLAs and the mass-metallicity relation for galaxies, Khare et al. (2007) suggested that sub-DLAs may arise in more massive galaxies, on average, than DLAs. 
This suggestion is  not that shocking, once one realizes that a higher H I column density by no 
means implies a higher total mass. Indeed, the suggestion seems reasonable given that sub-DLAs have somewhat 
higher velocity spreads on average than DLAs, and that a higher fraction of sub-DLAs have
 $\Delta v > 150$ km s$^{-1}$  (see Table 3). (Of course, as stressed by Kulkarni et al. 2007, not every sub-DLA is metal-rich. 
Thus, it is not necessary that every sub-DLA should be massive and have a large velocity width.)

To further explore whether sub-DLAs can arise in more massive galaxies than DLAs,  we now compare the metallicity evolution of DLAs and sub-DLAs with galaxy observations.  Our use of the robust metallicity indicators Zn and S 
enables us to make meaningful comparisons with galaxy data. We show that 
the  possibility of sub-DLAs being more massive than DLAs is consistent with stellar [Mn/Fe] ratios in 
galaxies, with galaxy down-sizing, with evolution of luminous star forming galaxies, and with galaxy number densities. 

{\bf Comparison with [Mn/Fe] ratios in galaxies:} 
In the stars of the Milky Way, [Mn/Fe] is found to increase with increasing [Fe/H], 
which could indicate production of Mn in type Ia supernovae (e.g., Samland 1998), or  metallicity-dependent yields in type II supernovae 
(e.g., Timmes, Woosley, \& Weaver 1995; McWilliam et al 2003; Feltzing et al. 2007).  
McWilliam et al. (2003) found that the [Mn/Fe] vs. [Fe/H] trend for 
the Sagittarius dwarf spheroidal is lower than that for the Galactic bulge stars by 
$\sim 0.2$ dex. They interpreted this difference to be a result of less rapid chemical evolution of 
the Sagittarius dwarf compared to the Galactic bulge: the Galactic bulge experienced rapid enrichment with 
time scales of $ < 1 $ Gyr (e.g., Matteucci, Romano, \& Molaro 1999), while dwarf spheroidal galaxies have extended star formation histories lasting many Gyrs 
(e.g, Mateo 1998; Grebel 2000). 

DLAs and sub-DLAs also 
show increasing [Mn/Fe] with increasing [Zn/H]. However, the two trends differ, suggesting different stellar populations.  Based on our measurements from 
Meiring et al. (2009a), we find that the linear regression slopes of the [Mn/Fe] vs. [Zn/H] trends are $0.27 \pm 0.03$ and $0.06 \pm 0.03$, respectively, for sub-DLAs and DLAs. The slopes of the corresponding relations for the Galactic bulge stars in McWilliam et al. (2003) and the thick disk stars in Reddy et al. (2006) are $0.40 \pm 0.05$ and 
$0.34 \pm 0.02$ respectively. The slope for the stars in the Sagittarius dwarf spheroidal galaxy in McWilliam et al. (2003) is $0.15 \pm 0.04$. The [Mn/Fe] vs. [Zn/H] trend in sub-DLAs appears to be closer to that for  bulge stars of the Milky Way, while that in DLAs 
is closer to the trend for the Sagittarius dwarf. This is consistent with the idea that sub-DLAs may have an older stellar population that got enriched faster than the stellar population in DLAs. We caution, however, that the existing data on [Mn/Fe] in galaxies is limited, and dwarf galaxies can have a range of Mn/Fe values depending on their star formation history (e.g., Cescutti et al. 2008).

{\bf Consistency with galaxy down-sizing:} 
The star formation history of galaxies shows evidence of Òdown-sizingÓ: the more massive galaxies experience ÒquenchingÓ of star formation activity at earlier times and get enriched faster than less massive galaxies (e.g., Cowie et al. 1996; Bundy et al. 2006).  If sub-DLAs arise in more massive galaxies than DLAs, then the faster metallicity evolution of sub-DLAs would be naturally expected from down-sizing. 

Furthermore, the existing sub-DLA metallicity data pose no serious conflict with 
the observations of massive galaxies being already metal-rich at $z \sim 2$ (e.g., Panter et al. 2008), for the following reasons. 

1. Fig. 1 shows that within the error bars, the mean sub-DLA metallicity at $z \sim 2$ is consistent with being only 0.6 dex below the solar value. Thus, there is no evidence that sub-DLAs evolved much 
later in metallicity compared to massive galaxies. (A suggestion to the contrary in some earlier works based on Fe/H vs. $z$ does not account for the fact that the Fe that is claimed to be too low could be locked up in dust grains.) Of course,  data on Zn and S in a larger sample of sub-DLAs would help to address this issue further. 

2.  Fig. 3 of Panter et al. (2008)  shows the metallicity-redshift relation for gas in star-forming galaxies. 
 Comparing this figure with our Fig. 1, the sub-DLA data at $z \gtrsim 1.1$ lie between the green and purple curves in the former figure (stellar mass-ranges of 1-3 $\times 10^{10} M_{\odot}$ and $ < 1 \times 10^{10} M_{\odot}$, respectively). In other words, existing sub-DLA data are consistent with their being 
moderately massive galaxies. 

3. Erb et al. (2006) showed that there is an offset in the mass-metalicity relations at $z\sim 0$ and $z \sim 2$, consistent with our metallicity evolution results, within the error bars. 

As noted before, the sub DLAs may not be the most massive galaxies. In fact, massive ellipticals may have such hot gas that the absorbers we normally know of evaporate, heated by infall; in this case we would not sample such galaxies even though they may have super-solar abundances. 

{\bf Comparison with metallicity evolution of luminous 
star-forming galaxies}:  
It has been pointed out that star forming galaxies (SFGs) show a shallower rise in metalicity with decreasing redshift than do sub-DLAs. However, the error on the slope of the SFG relationship is not well enough determined to know if that is true or not.  The slope of the metallicity evolution for SFGs is -0.15  according to Li (2008) , 
but $\approx -0.3$ at $z < 0.4$ according to Lara-L\'opez et al. (2009).  Thus, it is not clear 
whether there is a significant difference from the slope for sub-DLAs  ($-0.46 \pm 0.18$).  

Furthermore, there is no a-priori reason to expect the chemical evolution of sub-DLAs to be similar to 
that of luminous star-forming galaxies. Massive (or moderately massive) galaxies are not always 
forming stars.  Sub-DLAs could well 
have already passed through the star formation stage (as expected in the down-sizing 
scenario if they are more massive). 
Indeed, the few attempts made so far to detect 
direct signatures of star formation in absorbers at $z \gtrsim  0.7$ have often led to non-detections or weak detections of emission lines (Ly-$\alpha$, [O~II], H-$\alpha$, etc.), 
implying relatively low star formation 
rates (SFRs; e.g., Kulkarni et al. 2000, 2001, 2006, P\'eroux et al. 2010a, and references therein; Rauch  et al. 2008; 
Straka et al. 2010). 

{\bf Consistency with Number Densities:}  
The natural, but unfounded, association of high H I column density of a single absorbing cloud with the entire mass of a galaxy might make the number ratios of sub-DLAs to DLAs (a factor of $\sim 4$) seem to contradict the possibility that sub-DLAs come from more massive galaxies. We point out, however, that the number density of absorption lines per unit redshift ($dN/dz$) depends on the 
product of the cross-section of the absorbers, and 
their number density per unit co-moving volume (see, e.g., Sargent et al. 1980). Even though the number density of massive galaxies per unit co-moving 
volume is smaller than that of less massive ones, the cross-section for absorption of massive galaxies is higher than that of less massive ones. For example, the  
slope of the galaxy stellar mass function is  $\sim -0.4$ in the range 
$8 \lesssim$ log $(M_{*}/M_{\odot}) \lesssim 10.3$ (e.g., Baldry, Glazebrook, \& Driver 2008).  But the cross-section rises more rapidly with mass. Using hierarchical galaxy formation models, Okoshi \& Nagashima (2005) show that the average cross-section of an absorber is expected to 
scale as $\sigma \propto M_{\rm H I}^{0.97}$ for $6.5 \lesssim$ log $(M_{\rm H I} /M_{\odot}) \lesssim  10.5$, which also agrees with observations of local H I-rich galaxies (e.g., Rosenberg \& Schneider 2003). 
Therefore, the 
product of galaxy number density and cross-section, and hence $dN/dz$, should increase 
with the mass. Thus, if sub-DLAs are more massive than DLAs, then the fact that sub-DLAs outnumber DLAs could be a result of the larger gas cross-section for sub-DLAs and can be consistent with the relative fractions of more massive vs. less massive galaxies. Indeed, far more sight-lines through the Milky Way would be 
considered sub-DLAs, than those that would qualify as DLAs. 
 
\section{Conclusions and Future Work}
\label{6}

We have examined the metallicity evolution of DLAs and sub-DLAs using the largest sample 
yet of the nearly undepleted elements Zn and S, and compared them to properties of galaxies. Our main conclusions are as follows: \\
(1)  The higher mean metallicities of sub-DLAs compared to DLAs appear to be real, and not 
an artefact of Mg II selection, Fe II selection, or other systematic effects. \\
(2) The  intercept of the metallicity-redshift relation based on Zn or S is higher than that based on Fe by 0.5-0.6 dex for both DLAs and sub-DLAs. \\
(3) A correlation of metallicity with $W_{2796}$ or $W_{2599}$ is seen for DLAs, but not for sub-DLAs. \\
(4) The sub-DLAs have somewhat higher velocity spreads than DLAs, and a higher fraction of systems with $\Delta v > 150 $ km s$^{-1}$. \\
(5) Compared to DLAs, the sub-DLAs show a steeper growth of 
[Mn/Fe] with increasing metallicity that is closer to the trend for Galactic bulge/ thick disk stars. \\
(6) The absorber data are consistent with galaxy down-sizing. \\
(7) The data are also consistent with the relative number densities of low-mass and high-mass galaxies. \\
(8) It is plausible that sub-DLAs arise in more massive galaxies on average than DLAs.

Why is there a metallicity difference between DLAs and sub-DLAs? 
We suggest the physical reason for this difference is that they differ in their star formation histories. Sub-DLAs, being more massive on average than DLAs, underwent more rapid star formation and 
gas consumption, leaving them with lower $N_{\rm H I}$ and higher average metallicity--as also expected from galaxy down-sizing. The difference between sub-DLAs and DLAs may 
be qualitatively similar to the dichotomy between early-type and late-type galaxies.

We stress that a commonly hypothesized picture associating absorbers of decreasing $N_{\rm H I}$ 
with progressively 
larger  regions around galaxies does not agree with the observed metallicity differences between DLAs and sub-DLAs. 
If sub-DLAs were sampling outer parts of a galaxy, one would expect them to be less metal-rich than DLAs, in contrast to the observations. In the scenario we suggest, sub-DLAs have lower $N_{\rm H I}$ not because they sample outer parts of galaxies, but 
 because they experience more rapid star formation that leaves less neutral gas remaining. We also emphasize that there is little evidence for the hypothesis associating 
 decreasing $N_{\rm H I}$ with larger regions. The existing spectroscopically confirmed detections of galaxies producing DLAs/sub-DLAs show no systematic increase in impact parameter with decreasing $N_{\rm H I}$ (e.g., P\'eroux et al. 2010a; Meiring et al. 2010). Another way to test this scenario would be by observing DLAs/sub-DLAs toward gravitationally lensed or close binary quasars that sample multiple sight-lines at different impact parameters through foreground galaxies.

Some hints that metal-rich absorbers may arise in massive galaxies are seen in the small amount of imaging data available so far. A recent adaptive optics imaging study suggests a massive bulge-dominated galaxy as the source of 
a supersolar absorber at $z = 0.7$ (Chun et al. 2010). Furthermore, based on kinematics of H-$\alpha$ emission lines detected in VLT SINFONI integral field spectroscopy for two quasar absorbers (one DLA and one sub-DLA), 
P\'eroux et al. (2010b) estimate the halo mass of the supersolar sub-DLA galaxy to be $\sim 25$ times that of the 
relatively metal-poor DLA galaxy. 
Some other recent imaging studies have found multiple candidate galaxies near metal-rich sub-DLAs  (e.g., Hewett \& Wild 2007; Straka et al. 2010), suggesting that these metal-rich sub-DLAs may be associated with galaxy groups or clusters. (We note, however, that in low-$z$ galaxies, metallicity 
shows only a modest correlation with environment [e.g., Cooper et al. 2008; Ellison et al. 2009b]).

 We also note the possibility that the H I column density of a given absorber may change 
with time, as a result of the interaction between stars and interstellar gas. For example, star formation and/or gas stripping may cause a DLA to lose some H I and turn into a less gas-rich but more metal-rich sub-DLA. On the other hand, mergers may cause sub-DLAs poorer in H I to evolve into more gas-rich DLAs. 
Thus, the DLA and sub-DLA populations may not always be 
distinct. 

Further studies of the number density distribution per unit redshift per unit H I column density interval $d^{2}N / dz dN_{\rm H I}$ at low redshift are essential for both 
DLAs and sub-DLAs, to help understand whether a causal connection could exist between DLA and sub-DLA 
populations. Future high S/N UV spectroscopy of quasars with low-$z$ sub-DLAs  will help toward a better understanding of $d^{2}N / dz dN_{\rm H I}$ by reducing the incompleteness 
at the low-$N_{\rm H I}$ end.  Larger abundance samples for nearly undepleted elements 
like Zn and S are essential to better constrain the evolution of sub-DLAs and DLAs. Relative abundances 
such as [Mn/Fe] in a large sample would help to better understand the stellar populations in DLAs and sub-DLAs. Finally, it is essential to image the galaxies producing the absorbers 
and measure their masses, morphologies, and SFRs to directly understand how DLA and sub-DLA  populations 
differ and how they are related to each other.  

\setlength{\tabcolsep}{5pt}
\begin{table}[h]
\begin{center}
\caption{$W_{2796}$-metallicity Correlations for DLAs (D) and Sub-DLAs (S). The statistic used 
is the generalized Kendall's $\tau$ correlation coefficient with survival analysis. \label{tbl-1}}
\begin{tabular}{lll}
\hline\hline
Type & Quantity &  Result
 \\
\hline
S& [X/H]$^{a}$, $W_{2796}$ & $\tau = -0.070$, $P = 0.670$\\
S & [Fe/H], $W_{2796}$ &   $\tau = 0.128$, $P = 0.541$\\
D& [X/H]$^{a}$, $W_{2796}$ &  $\tau = 0.819$, $P < 0.0001$\\
D&[Fe/H], $W_{2796}$ &   $\tau = 0.752$, $P = 0.010$\\
\hline
\end{tabular}
\begin{minipage}{\linewidth}
$^{a}${X = Zn or S}
\end{minipage}
\end{center}
\end{table}

\setlength{\tabcolsep}{5pt}
\begin{table}[h]
\begin{center}
\caption{$W_{2599}$-metallicity Correlations for DLAs (D) and Sub-DLAs (S). The statistic used 
is the generalized Kendall's $\tau$ correlation coefficient with survival analysis. \label{tbl-2}}
\begin{tabular}{lll}
\hline\hline
Type & Quantity &  Result
 \\
\hline
S& [X/H]$^{a}$, $W_{2599}$ & $\tau = 0.171$, $P = 0.302$\\
D& [X/H]$^{a}$, $W_{2599}$ &  $\tau = 0.744$, $P = 0.0005$\\
\hline
\end{tabular}
\begin{minipage}{\linewidth}
$^{a}${X = Zn or S}
\end{minipage}
\end{center}
\end{table}

\section*{Acknowledgments}
VPK and DS gratefully acknowledge partial support from the US National Science Foundation grants  AST-0607739 and AST-0908890 (PI Kulkarni). We thank an anonymous referee for helpful comments. We also thank S. Ellison and M. Dessauges-Zavadsky for comments on an earlier version of this paper. 

\vskip 0.5in

\begin{table}[h]
\begin{center}
\caption{Velocity Spread Distributions of DLAs (D) and Sub-DLAs (S) \label{tbl-3}}
\begin{tabular}{llll}
\hline\hline
Type & Quantity & Statistic  &  Result
 \\
\hline
S & $\Delta v$ (weak lines)&Mean, Median & 125.5, 95.3 km s$^{-1}$ \\
D & $\Delta v$ (weak lines)&  Mean, Median &  103.0, 80.0 km s$^{-1}$ \\
S  & $\Delta v$ (Mg II 2796)&  Mean, Median & 225.0, 197.0 km s$^{-1}$ \\
D  & $\Delta v$ (Mg II 2796)& Mean, Median & 191.8, 177.0 km s$^{-1}$ \\ 
S & $\Delta v$ (weak lines) &  Fraction with & $ 30.4 \pm 8.1\%$ \\
&&$\Delta v > 150$&\\
S & $\Delta v$ (weak lines) & Fraction with & $ 17.4 \pm 6.1\%$ \\
&&$\Delta v < 50$&\\
D & $\Delta v$ (weak lines) &Fraction with & $ 11.9 \pm 4.2\%$ \\
&&$\Delta v > 150$&\\
D & $\Delta v$ (weak lines) &Fraction with & $ 29.9 \pm 6.7\%$ \\
&&$\Delta v < 50$&\\
\hline
\end{tabular}
\end{center}
\end{table}

\pagebreak

\begin{table}[h]
\begin{center}
\caption{Velocity-Metallicity Correlations for DLAs (D) and Sub-DLAs (S). The statistic used 
is the generalized Kendall's $\tau$ correlation coefficient with survival analysis. \label{tbl-4}}
\begin{tabular}{lll}
\hline\hline
Type & Quantity & Result
 \\
\hline
D &[X/H]$^{a}$, $\Delta v$ (weak)& $\tau = 0.914$,\\
&&$P < 0.0001$\\
S &[X/H]$^{a}$, $\Delta v$ (weak)& $\tau = 0.106$,\\
&&$P=0.526$\\
D&[Fe/H], $\Delta v$ (weak)&  $\tau = 0.960$, \\
&&$P < 0.0001$\\
S&[Fe/H], $\Delta v$ (weak)&   $\tau = 0.498$,\\
&&$P=0.0277$\\ 
D&[X/H], $\Delta v$ (Mg II)&   $\tau = 0.765$,\\
&&$P=0.017$\\
S &[X/H], $\Delta v$ (Mg II)&   $\tau = -0.097$,\\
&&$P=0.638$\\
\hline
\end{tabular}
\begin{minipage}{\linewidth}
$^{a}${X = Zn or S}
\end{minipage}
\end{center}
\end{table}

\vskip 0.5in

\begin{figure}[h]
\begin{center}
\includegraphics[angle=0, width=\linewidth]{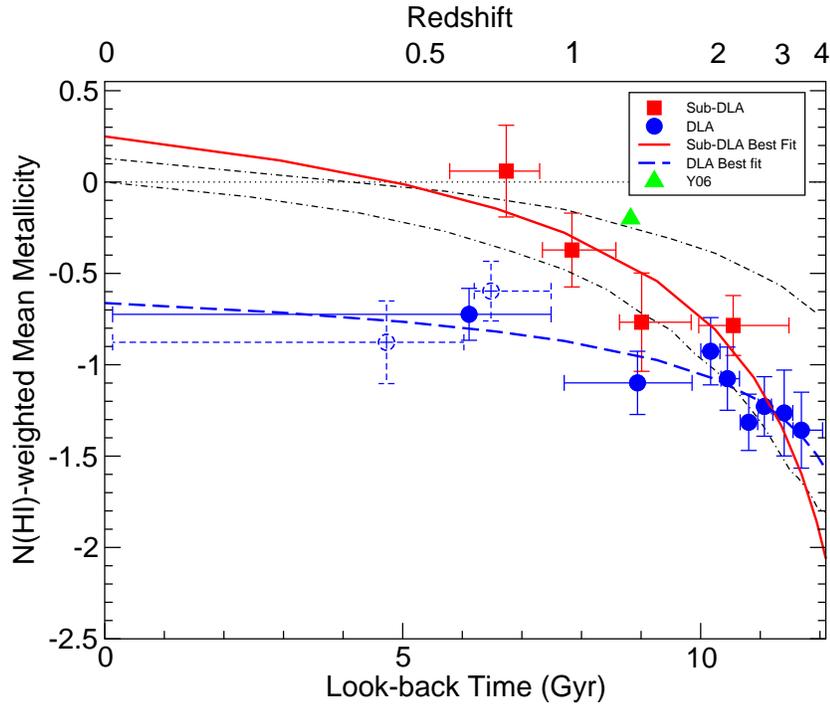}
\caption[Fig1]{  $N({\rm H \, I})$-weighted mean  
metallicity vs. look-back time relation for 154  
DLAs and 58 sub-DLAs with Zn or S measurements. Filled circles show 8 bins with 19 or 20 DLAs each. 
Unfilled circles refer to the lowest time bin for DLAs split into 2 bins with 10 
DLAs each. Squares denote 4 bins with 14 or 15 sub-DLAs each. Horizontal bars denote ranges in look-back times 
covered by each bin. Vertical errorbars denote 1 $\sigma$ uncertainties. The triangle shows the formal lower limit to the average metallicity 
for a composite spectrum from 698 
absorbers with average log $N_{\rm H \, I} \sim 20$ (sample 24) 
from York et al. (2006). The bold solid and dashed curves show the best fits obtained from linear regression of the metallicity vs. redshift data for sub-DLAs and DLAs, respectively. 
The light dot-dashed and dot-double-dashed curves show, 
respectively, the mean metallicity in the models 
of Pei et al. (1999) and Somerville et al. (2001). Sub-DLAs appear to be more metal-rich and faster-evolving than DLAs, especially at lower redshifts. \label{fig1}}
\end{center}
\end{figure}			
\vskip -0.5in

\begin{figure}[h]
\includegraphics[angle=0, width=4.5in]{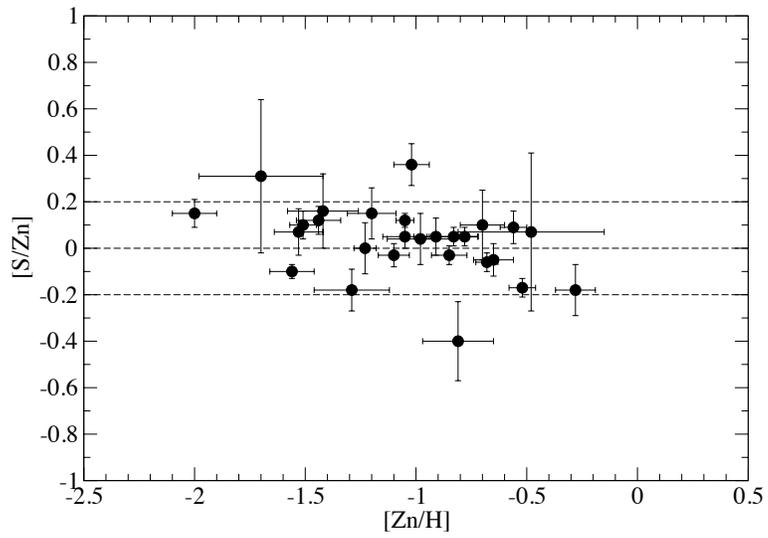}
\caption{[S/Zn] vs. [Zn/H] for absorbers with detections of S and Zn. Dashed horizontal lines denote the solar level and $\pm 0.2$ dex levels. S/Zn tracks the 
solar ratio within $\pm 0.2$ dex for most of the absorbers. \label{fig2}}
\end{figure}

\begin{figure}[h]
\includegraphics[angle=0, width=4.5in]{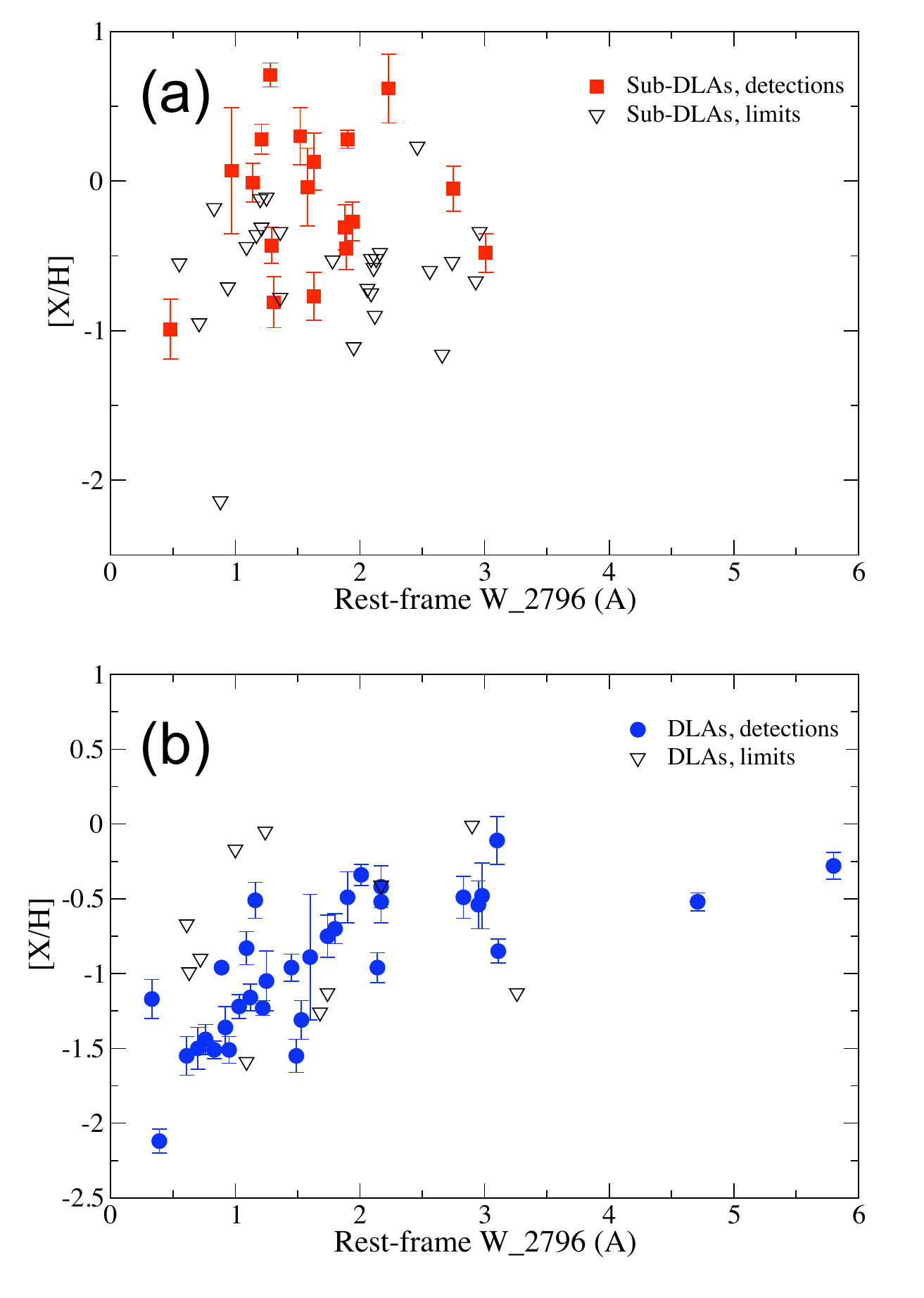}
\caption{Rest-frame equivalent width of Mg II $\lambda 2796$ plotted vs. metallicity [X/H] from Zn or S, for (a) sub-DLAs ($\it {top}$) and (b) DLAs ($\it{bottom}$). No correlation is seen for sub-DLAs, whereas a correlation is seen for DLAs. Thus sub-DLA data do not have an Mg II-selection bias toward high-metallicities. \label{fig3}}
\end{figure}

\begin{figure}[h]
  {\begin{minipage}[t]{150pt}
  \vskip -0.35in
  \hskip -0.35in
    \includegraphics[angle=0, width=4.0in]{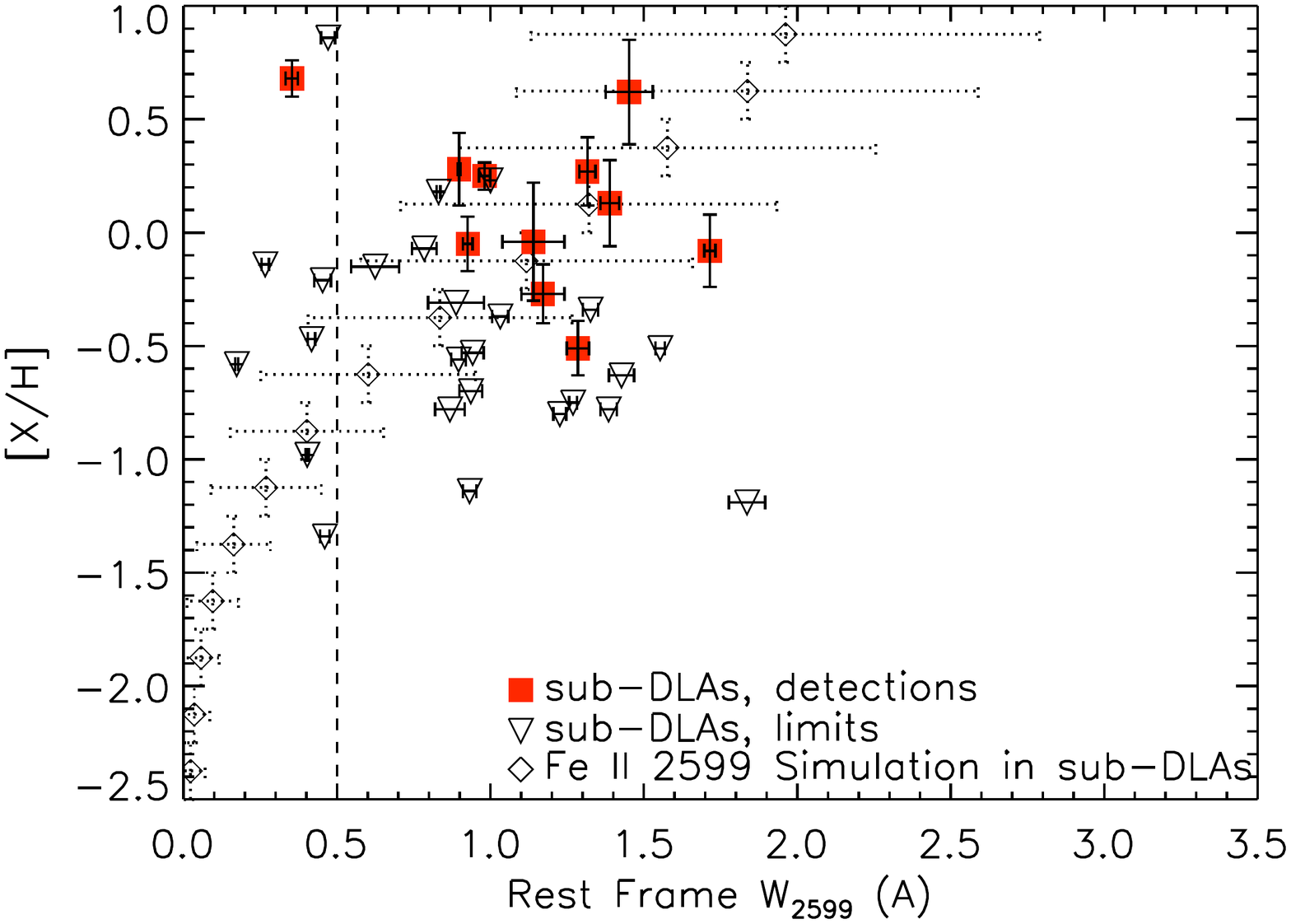}
  \end{minipage}}
  \vskip -0.5in
  {\begin{minipage}[t]{150pt}
  \vskip -0.15in
  \hskip -0.35in
    \includegraphics[angle=0, width=4.0in]{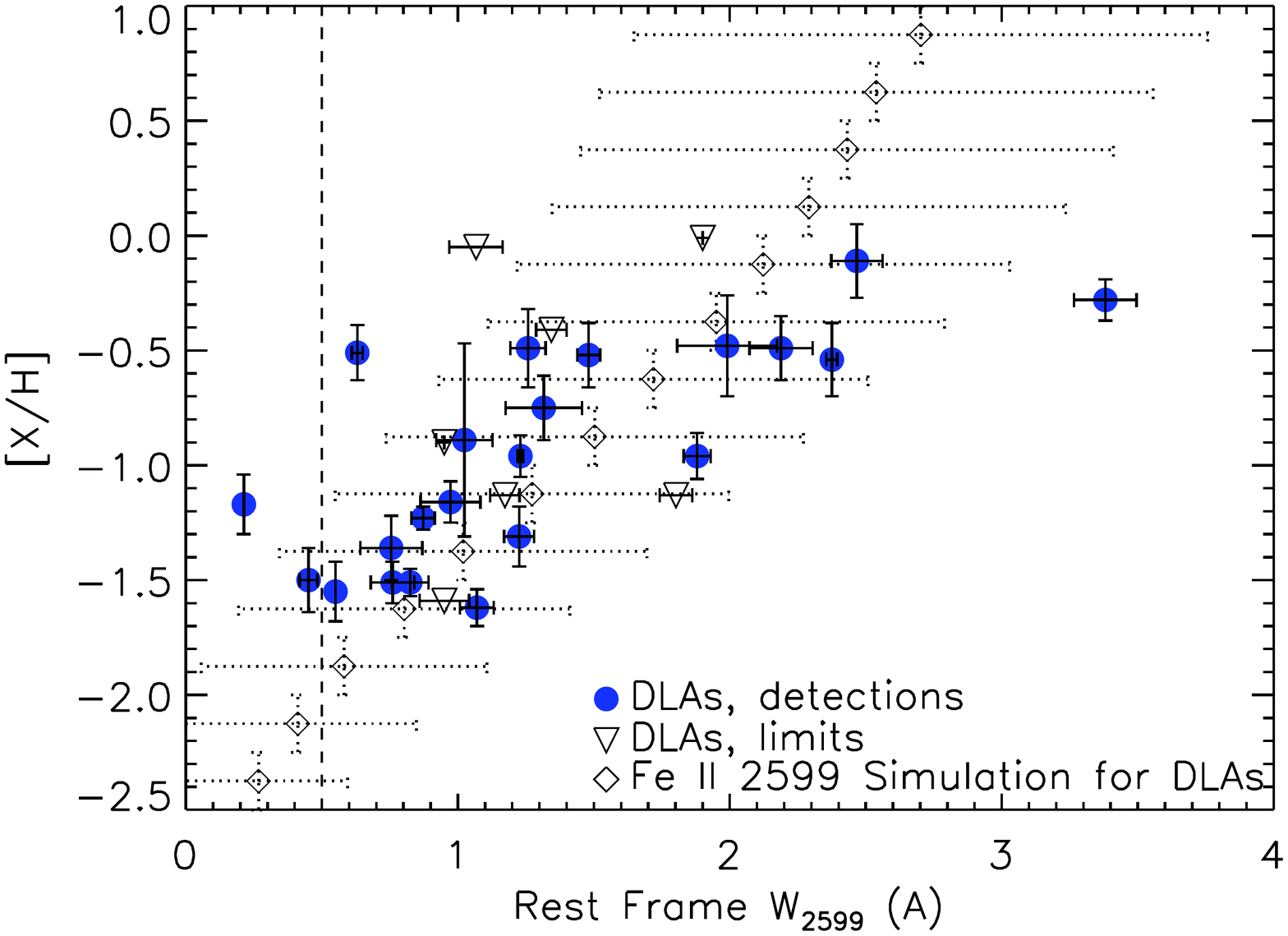}
  \end{minipage}}
  \vskip -0.2in
  \caption{Comparison of data and simulations for metallicity vs. rest-frame equivalent width of Fe II $\lambda 2599$, for (a) sub-DLAs ($\it {top}$), and (b) DLAs ($\it{bottom}$). The squares, triangles, and circles show $W_{2599}$ plotted vs. metallicity [X/H] for the absorbers in our sample where $W_{2599}$ data are available.  The diamonds show median $W_{2599}$ values from 30,000 simulations, plotted as a function of metallicity,  in bins of 0.25 dex in [X/H]. The dashed horizontal error bars on the diamonds show the  $\pm 1 \, \sigma$ ranges in the $W_{2599}$ values generated in the simulations. The DLA data show a correlation 
 between the metallicity and $W_{2599}$, but the sub-DLA 
 data do not show a correlation. Thus, sub-DLA data do not have an Fe II selection bias toward high 
 metallicities. \label{fig4}}
\end{figure}

\begin{figure}[h]
 {\begin{minipage}[t]{150pt}
  \vskip -0.35in
  \hskip -0.35in
    \includegraphics[angle=0, width=4.0in]{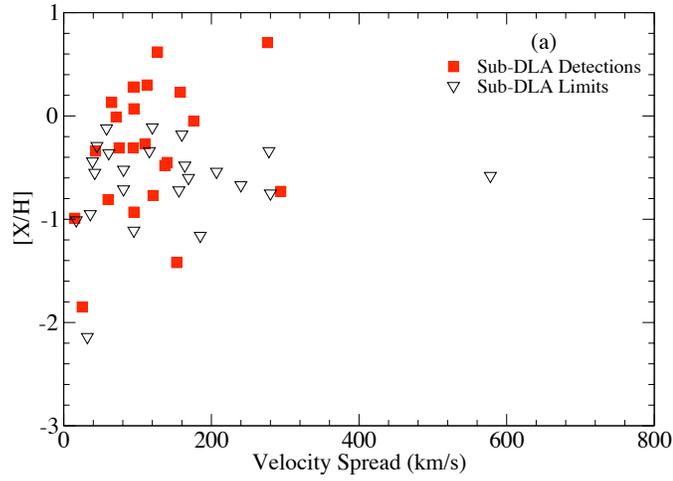}
  \end{minipage}}
  \vskip 0.0in
  {\begin{minipage}[t]{150pt}
  \vskip -0.15in
  \hskip -0.35in
    \includegraphics[angle=0, width=4.0in]{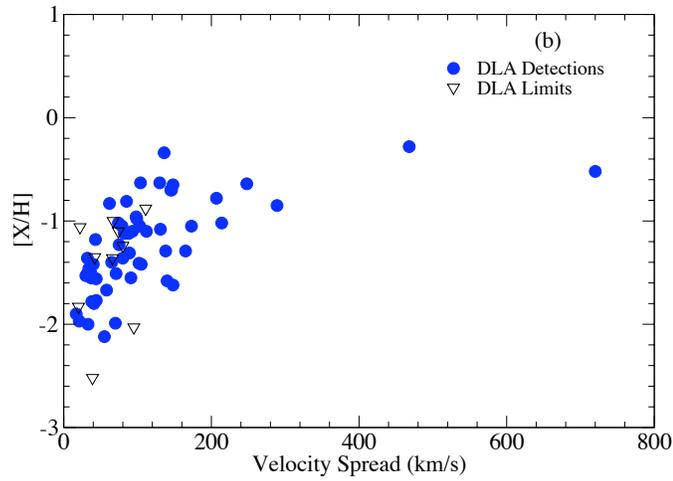}
  \end{minipage}}
  \vskip -0.2in
\caption{Velocity spread $\Delta v$, determined from weak absorption lines, plotted vs. metallicity [X/H] from Zn or S, for (a) sub-DLAs  and (b) DLAs. The correlation is stronger for DLAs than for sub-DLAs.  \label{fig5}}
\end{figure}

\bibliographystyle{elsarticle-harv}
\bibliography{<your-bib-database>}

\begin{thebibliography}{99}



\bibitem[Baldry et al.(2008)]{Baldry08} Baldry I. K., Glazebrook K., Driver S. P., 2008, MNRAS, 388, 945

\bibitem[Bundy et al.(2006)]{Bundy06} Bundy K. et al., 2006, ApJ, 651, 120

\bibitem[Cescutti et al.(2008)]{Cesutti08} Cescutti G., Matteuci F., Lanfranchi, G. A., McWilliam, A., 2008, A\&A, 491, 401

\bibitem[Chen et al.(2005)]{Chen05} Chen H.-W., Kennicutt R. C., Rauch M., 2005, ApJ, 620, 703 

\bibitem[Chun et al.(2009)]{Chun09} Chun M. R., Kulkarni V. P., Gharanfoli S., Takamiya M., 2010, 
AJ, 139, 296

\bibitem[Cooper et al.(2008)]{Cooper08} Cooper M. C., Tremonti C. A., Newman J. A.,  Zabludoff A. I., 2008, 
MNRAS, 390, 245 

\bibitem[Cowie et al.(1996)]{Cowie96} Cowie L. L., Songaila A., Hu E. M., Cohen J. G., 1996, AJ, 112, 839

\bibitem[Dessauges-Zavadsky et al.(2009)]{D09} Dessauges-Zavadsky M., Ellison S.  L, Murphy M. T., 2009, MNRAS, 396, L61 

\bibitem[Ding et al.(2003)]{Ding03} Ding J., Charlton J. C., Churchill C. W., Palma C., 2003, ApJ, 590, 746

\bibitem[Ellison et al.(2005)]{Ell05} Ellison S., Kewley L. J., Mallen-Omelas G., 2005, MNRAS, 357, 354

\bibitem[Ellison et al.(2009a)]{Ell09a} Ellison S. L., Murphy M. T., Dessauges-Zavadsky M., 2009a, MNRAS, 392, 998

\bibitem[Ellison et al.(2009b)]{Ell09b} Ellison S. L., Simard L., Cowan N. B., Baldry I. K., Patton D. R., McConnachie A. W., 2009b, MNRAS, 396, 1257

\bibitem[Erb et al.(2006)]{Erb06} Erb D. K., Shapley A. E., Pettini M., Steidel C. C., 
Reddy N. A.,  Adelberger K. L., 2006,  ApJ, 644, 813

\bibitem[Feltzing et al.(2007)]{Felt07} Feltzing S., Fohlman M., Bensby T., 2007, A\&A, 467, 665

\bibitem[Ferguson et al.(1998)]{Ferg98} Ferguson A. M. N., Gallgher J. S., Wyse R. F. G., 1998, AJ, 116, 673

\bibitem[Grebel et al.(2000)]{Gre00} Grebel, E. K. 2000,  in 33d ESLAB Symp., Star Formation from the Small to 
the Large Scale, ed. F. Favata, A. A. Kaas, \& A. Wilson (ESA SP-445; 
Noordwijk: ESA), 87

\bibitem[Herbert-Fort et al.(2006)]{HF06} Herbert-Fort S., Prochaska J. X., Dessauges-Zavadsky M., Ellison S. L., Howk J. C., Wolfe A. M., Prochter G. E., 2006, PASP, 118, 1077

\bibitem[Hewett et al.(2007)]{HW07} Hewett P.-C., Wild V., 2007, MNRAS, 379, 738

\bibitem[Ivezi et al.(2008)]{Ivezic08} Ivezi\' c Z., Sesar B., Juric M. et al., 2008, ApJ, 684, 287 

\bibitem[Jenkins et al.(2005)]{Jen05} Jenkins E. B., Bowen D. V., Tripp T. M., Sembach K. R., 2005, ApJ, 623, 767

\bibitem[Jenkins et al.(2009]{Jen09} Jenkins E. B., 2009, ApJ, 700, 1299

\bibitem[Khare et al.(2007)]{Kh07} Khare P., Kulkarni V. P., P\'eroux C., York D. G., Lauroesch J. T., Meiring J. D., 2007,  A\&A, 464, 487 

\bibitem[Kulkarni et al.(2000)]{K00} Kulkarni V. P., Hill J. M.,  Schneider G., Weymann R. J., Storrie-Lombardi L. J., Rieke M. J., Thompson R. I., Jannuzi B., 2000, ApJ, 536, 36

\bibitem[Kulkarni et al.(2001)]{K01} Kulkarni V. P., Hill J. M., Schneider G., Weymann R. J., Storrie-Lombardi L. J., Rieke M. J., Thompson R. I., Jannuzi B., 2001, ApJ, 551, 37

\bibitem[Kulkarni et al.(2002)]{K02} Kulkarni V. P., Fall S. M., 2002, ApJ, 580, 732 

\bibitem[Kulkarni et al.(2005)]{K05} Kulkarni V. P., Fall S. M., Lauroesch J. T., York D. G., Welty D. E., Khare P., Truran J. W., 2005, ApJ, 618, 68

\bibitem[Kulkarni et al.(2006)]{K06} Kulkarni V. P., Woodgate B. E., York D. G., Thatte D. G., Meiring J., Palunas P., Wassell E., 2006, 
ApJ, 636, 30 

\bibitem[Kulkarni et al.(2007)]{K07} Kulkarni V. P., Khare P., P\'eroux C., York D. G., Lauroesch J. T.,  Meiring J. D.,  2007, ApJ, 661, 88

\bibitem[Lara-Lopez et al.(2009)]{LL09} Lara-L\'opez M. A., Cepa J.,  Bongiovanni A., P\'erez Garc'a A. M., Castaneda H.,  Fernandez L. M., Povic M.,  Sanchez-Portal M., 2009, A\&A, 505, 529

\bibitem[Ledoux et al.(2003)]{Led03} Ledoux C., Petitjean P., Srianand R., 2003, MNAS, 346, 209

\bibitem[Ledoux et al.(2006)]{Led06} Ledoux C., Petitjean P., Fynbo J. P. U., Moller P.,  Srianand R., 2006,  A\&A, 457, 71

\bibitem[Lemasle et al.(2008)]{Lem08} Lemasle B., Francois P., Piersimoni A., Pedicelli S., Bono G., Laney C. D., Primas F., Romaniello M.,  2008, A\&A, 490, 613 

\bibitem[Li et al.(2008)]{Li08} Li L.-X., 2008, MNRAS, 388, 1487

\bibitem[Maseiro et al.(2005)]{Mas05} Maseiro J. D. R., Charlton J. C., Ding J., Churchill C. W., Kacprzak G., 2005, 
ApJ, 623, 57

\bibitem[Mateo et al.(1998)]{Mat98} Mateo, M. 1998, ARAA, 36, 435

\bibitem[Matteucci et al.(1999)]{Mat99} Matteucci, F., Romano, D., Molaro, P. 1999, A\&A, 341, 458 

\bibitem[McWilliam et al.(2003)]{MacW03} McWilliam A., Rich R. M., Smecker-Hane T. A., 2003, ApJ, 592, L21 
  
\bibitem[Meiring et al.(2006)]{Mei06} Meiring J. D., Kulkarni V. P., Khare P., Bechtold J., 
York D. G., Cui J., Lauroesch J. T., Crotts A. P. S., Nakamura O., 2006, MNRAS, 370, 43

\bibitem[Meiring et al.(2007)]{Mei07} Meiring J. D., Lauroesch J. T., Kulkarni V. P., P\'eroux C., Khare P., York D. G., Crotts A. P. S.,  2007,  MNRAS, 376, 557

\bibitem[Meiring et al.(2008)]{Mei08} Meiring J. D., Kulkarni V. P., Lauroesch J. T., 
P\'eroux C., Khare P., York D. G., Crotts A. P. S.,  2008,  
MNRAS, 384, 1015

\bibitem[Meiring et al.(2009a)]{Mei09a} Meiring J. D., Kulkarni V. P., Lauroesch J. T., 
P\'eroux C., Khare P., York D. G., Crotts A. P. S.,  2009a,  
MNRAS, 393, 1513

\bibitem[Meiring et al.(2009b)]{Mei09b} Meiring J. D., Lauroesch J. T., Kulkarni V. P., 
P\'eroux C., Khare P., York D. G., Crotts A. P. S.,  2009b, MNRAS, 397, 2037 

\bibitem[Meiring et al.(2010)]{Mei10} Meiring, J., D., Lauroesch, J. T., Haberzettel, L. Kulkarni, V. P., 
P\'eroux, C., Khare, P., York, D. G. 2010, MNRAS, submitted

\bibitem[Miller et al.(2007)]{Mil07} Miller A.,  Lauroesch J. T., Sofia U. J.,  Cartledge S. I. B., Meyer D. M., 
2007, ApJ, 659, 441

\bibitem[Misawa et al.(2008)]{Mis08} Misawa T., Charlton J. C., Narayanan A., 2008, ApJ, 679, 220
 
\bibitem[Murphy et al.(2007)]{Mur07} Murphy, M. T., Curran, S. J., Webb, J. K., M\'enager, H., Zych, B. J., 2007, MNRAS, 376, 673

\bibitem[Nestor et al.(2008)]{Nest08} Nestor, D. B., Pettini, M., Hewett, P. C., Rao, S., Wild, V., 2008, MNRAS, 390, 1670

\bibitem[Nissen et al.(2004)]{Nis04} Nissen P. E., Chen Y. Q., Asplund M., Pettini M., 2004, A\&A, 415, 993

\bibitem[Nissen et al.(2007)]{Nis07} Nissen P. E., Akerman, C., Asplund M., Fabbian, D., Kerber, F., Kaufl, H. U., Pettini M., 2007, A\&A, 469, 319

\bibitem[Noterdaeme et al.(2008)]{Not08} Noterdaeme, P., Ledoux, C., Petitjean, P., Srianand, R., 2008, A\&A, 481, 327

\bibitem[Okoshi et al.(2005)]{ON05} Okoshi K., Nagashima M., 2005, ApJ, 623, 99

\bibitem[Panter et al.(2008)]{Pan08} Panter B. Jimenez R., Heavens A. F., Charlot S., 2008, MNRAS, 391, 1117

\bibitem[Pei et al.(1999)]{PFH99} Pei Y. C., Fall S. M., Hauser M. G., 1999,  ApJ, 522, 604 

\bibitem[P\'eroux et al.(2005)]{Per05} P\'eroux C.,Dessauges-Zavadsky M.; D'Odorico S.; Kim T.-S.; McMahon R. G., 2005, MNRAS, 363, 479

\bibitem[P\'eroux et al.(2006)]{Per06} P\'eroux C., Kulkarni V. P., Meiring J., Ferlet R., 
Khare P., Lauroesch J. T., Vladilo G., York D. G., 2006, A\&A, 
450, 53

\bibitem[P\'eroux et al.(2008)]{Per08} P\'eroux C., Meiring J. D., Kulkarni V. P., Khare P., Lauroesch J. T., Vladilo G., York D. 
G., 2008, MNRAS, 386, 2209

\bibitem[P\'eroux et al.(2010a)]{Per10} P\'eroux, C., Bouch\'e, N., Kulkarni, V. P., York, D. G.,  Vladilo, G., 2010a, MNRAS, submitted

\bibitem[P\'eroux et al.(2010b)]{Per10} P\'eroux, C., Bouch\'e, N., Kulkarni, V. P., York, D. G.,  Vladilo, G., 2010b, MNRAS, submitted


\bibitem[Prochaska et al.(2005)]{Pro05} Prochaska J. X., Herbert-Fort S., Wolfe A. M., 2005, ApJ, 635, 123

\bibitem[Prochaska et al.(2006)]{Pro06} Prochaska J. X., O'Meara J. M., Herbert-Fort S., Burles S., Prochter G. E., Bernstein R. A., 2006, ApJ, 648, L97

\bibitem[Quast et al.(2008)]{Quast08} Quast, R., Reimers, D., Baade, R., 2008, A\&A, 477, 443

\bibitem[Rauch et al.(2008)]{Rau08} Rauch M. et al., 2008, ApJ, 681, 856  

\bibitem[Reddy et al.(2006)]{Reddy06} Reddy B. E., Lambert D. L., Prieto C. A., 2006, MNRAS, 367, 1329

\bibitem[Rosenberg et al.(2003)]{RS03} Rosenberg J. L., Schneider S. E., 2003, ApJ, 585, 256

\bibitem[Routly et al.(1952)]{Rou52} Routly P. M., Spitzer L., 1952, ApJ, 115, 227

\bibitem[Samland et al.(1998)]{Sam98} Samland M., 1998, ApJ, 496, 155 

\bibitem[Sargent et al.(1980)]{Sar80} Sargent W. L. W., Young P. J., Boksenberg A., Tytler D., 1980, ApJSS, 42, 41

\bibitem[Savaglio et al.(2005)]{Sav05} Savaglio S. et al., 2005, ApJ, 635, 260

\bibitem[Shull et al.(2009)]{Shull09} Shull J. M., Jones J. R., Danforth C. W., Collins J. A., 2009, 
ApJ, 699, 754

\bibitem[Somerville et al.(2001)]{Som01} Somerville R. S., Primack J. R., Faber S. M., 2001, MNRAS, 320, 504

\bibitem[Srianand et al.(2008)]{Sri08} Srianand R., Noterdaeme P., Ledoux C., Petitjean P., 2008, A\&A, 482, L39

\bibitem[Storrie-Lombardi et al.(2000)]{SW00} Storrie-Lombardi L. J. Wolfe A. M., 2000, ApJ, 543, 552

\bibitem[Straka et al.(2009)]{Str09} Straka L. A., Kulkarni V. P., York D. G., Woodgate B. E., Grady C. A., 2010, AJ, 139, 1144

\bibitem[Timmes et al.(1995)]{TWW95} Timmes F. X., Woosley S. E., Weaver T. A., 1995, ApJS, 96, 617

\bibitem[Tremonti et al.(2004)]{Tre04} Tremonti C. A. et al., 2004, ApJ, 613, 898

\bibitem[Wakker et al.(2007)]{Wak07} Wakker B. P. et al., 2007, ApJ, 670, L113

\bibitem[York et al.(2006)]{Y06} York D. G. et al., 2006, MNRAS, 367, 945

\bibitem[Zwaan et al.(2005)]{Zwa05} Zwaan M. A., van der Hulst J. M., Briggs F. H., Verheijen  
M. A. W., Ryan-Weber E. V., 2005, MNRAS, 364, 1467 

 \end{thebibliography}



\end{document}